\documentclass[12pt]{article}
\usepackage{epsfig}

\usepackage{graphicx}
\usepackage{cancel}
\usepackage{amssymb,amsmath}

\def\harr#1#2{\smash{\mathop{\hbox to .3in{\rightarrowfill}}
 \limits^{\scriptstyle#1}_{\scriptstyle#2}}}

\def\s2{\frac{1}{\sqrt2}}

\def\be{\begin{equation}}
\def\ee{\end{equation}}
\def\beqa{\begin{eqnarray}}
\def\eeqa{\end{eqnarray}}
\def\tr{{\rm tr \,}}
\def\Tr{{\rm Tr \,}}

\def\Dsl{\,\raise.15ex\hbox{/}\mkern-13.5mu D} 
\def\d3{d^3}

\begin{document}

\begin{center}
\Large{\bf New Reflections on Gravitational Duality} \vspace{0.5cm}

\large Hugo Garc\'{\i}a-Compe\'an$^{a}$\footnote{e-mail address:
{\tt compean@fis.cinvestav.mx}}, Octavio
Obreg\'on$^{b}$\footnote{e-mail address: {\tt
octavio@fisica.ugto.mx}}, Cupatitzio Ramirez$^{c}$\footnote{e-mail
address: {\tt
cramirez@fcfm.buap.mx}}\\

\vspace{0.5cm} \vspace{0.5cm}
{\small \em $^a$Departamento de F\'{\i}sica, Centro de Investigaci\'on y de Estudios Avanzados del IPN}\\
{\small\em P.O. Box 14-740, CP. 07000, Ciudad de M\'exico., M\'exico.}\\
\vspace{0.5cm}
{\small \em $^b$Departamento de F\'{\i}sica, Divisi\'on de Ciencias e Ingenier\'{\i}a}\\
{\small\em Universidad de Guanajuato, Campus Le\'on}\\
{\small\em Loma del Bosque No. 103, Frac. Lomas del Campestre, Le\'on, Gto., M\'exico.}\\
\vspace{0.5cm} {\small \em $^c$Facultad de Ciencias F\'{\i}sico
Matem\'aticas, Benem\'erita Universidad
Aut\'onoma de Puebla}\\
{\small\em Puebla 72570, M\'exico.}\\
\vspace{0.5cm}

\vspace*{0.5cm}
\end{center}

\begin{abstract}
In general terms duality consists of two descriptions of one
physical system by using degrees of freedom of different nature.
There are different kinds of dualities and they have been extremely
useful to uncover the underlying strong coupling dynamics of gauge
theories in various dimensions and those of the diverse string
theories. Perhaps the oldest example exhibiting this property is
Maxwell theory, which interchanges electric and magnetic fields. An
extension of this duality involving the sources is also possible if
the magnetic monopole is incorporated. At the present time a lot has
been understood about duality in non-Abelian gauge theories as in
the case of ${\cal N}=4$ supersymmetric gauge theories in four
dimensions or in the Seiberg-Witten duality for ${\cal N}=2$
theories. Moreover, a duality that relates a gravitational theory
(or a string theory) and a conformal gauge theory, as in the case of
gauge/gravity correspondence, have been also studied with
considerable detail. The case of duality between two gravitational
theories is the so called {\it gravitational duality}. At the
present time, this duality has not been exhaustively studied,
however some advances have been reported in the literature. In the
present paper we give a general overview of this subject. In
particular we will focus on non-Abelian dualities, applied to
various theories of gravity as developed by the authors, based in
the Rocek-Verlinde duality procedure. Finally, as a new development
in this direction, we study the gravitational duality in Hitchin's
gravity in seven and six dimensions and their relation is also
discussed.

\vskip .3truecm

\noindent\leftline{August 22, 2018}
\end{abstract}

\bigskip
\newpage
\section{Introduction}
\label{sec:intro}

In general terms, {\it duality} consists of two descriptions of a
physical system through different degrees of freedom. There are
different kinds of dualities and they have been extremely useful to
uncover the underlying dynamics of strong coupling gauge theories in
various dimensions and those of the diverse string theories
\cite{Giveon:1998sr}.

In some cases this relation involves the inversion of the coupling
constant in such a way that non-perturbative phenomena of the
original degrees of freedom can be mapped to a perturbative theory
of the dual degrees of freedom. This duality is termed $S$-duality
and it will be the subject of the present article in the context of
gravitational theories. Here we will not intend to give an
exhaustive, complete and detailed overview on this subject, which is
out of our scope. Thus for a general review, see for instance
\cite{Nurmagambetov:2006vz,Dehouck:2011xt} and references therein.

The paradigmatic example satisfying the property of duality is the
Maxwell theory with magnetic charges and magnetic currents. At the
present time a lot has been understood about duality in gauge
theories. The most relevant examples are the ${\cal N}=4$
supersymmetric gauge theories in four dimensions, where the
Montonen-Olive duality \cite{Montonen:1977sn,Olive:1997fg} was
proved \cite{Vafa:1994tf}, and the Seiberg-Witten duality for ${\cal
N}=2$ theories (see for instance, \cite{AlvarezGaume:1996mv}).
Further, a duality that relates a gravitational theory (or a string
theory) and a conformal gauge theory, as in the case of
gauge/gravity correspondence
\cite{Maldacena:1997re,Gubser:1998bc,Witten:1998qj}, which has also
been studied with quite detail (for a survey see for instance,
\cite{Aharony:1999ti,Natsuume:2014sfa,Ammon:2015wua}). However, much
of the general work on duality has been done using the Feynman
functional integral and it has not been considered as a well
established mathematical result. This has motivated many
mathematicians and mathematical physicists to work on rigorous
proofs to support these results.

The case of duality between two gravitational theories is the so
called {\it gravitational duality}. At the present time this duality
has not been exhaustively understood and it is a conjectured
symmetry existing in some gravity theories or in the gravitational
sector of some higher-dimensional supergravity or superstring
theories; some advances have been reported in the literature
\cite{Nurmagambetov:2006vz,Dehouck:2011xt}. In fact, the reach of this analysis is
not comparable with those obtained in supersymmetric Yang-Mills
theories or superstring theories. In these latter theories the power
of strong/weak duality and T-duality in superstring theories has
allowed to compute many non-trivial observables carrying much
information on the system \cite{Giveon:1994fu}.

Very recently new advances in duality have been done in the context
of condensed matter systems, see for instance
\cite{Choudhury:2018iwf}. There is a conjecture asserting that
a fermionic system coupled to a Chern-Simons field
is dual to a Chern-Simons gauged Wilson-Fisher bosonic theory. This conjecture has been proved for the
case of negative mass deformation of the fermionic theory \cite{Choudhury:2018iwf}.

In \cite{Dijkgraaf:2004te} the so called {\it topological M-theory}
has been proposed. This is a gravity theory of three-forms on a
seven-dimensional manifold of $G_2$-holonomy, which has been
regarded as a master system from which it is possible to obtain, by
dimensional reduction, the different {\it form theories of gravity}
in lower dimensions, six, four and three. Among the known
theories of gravity are:  In six dimensions, on a Calabi-Yau manifold,
there are two theories of gravity describing the moduli of complex
structures and the K\"ahler cone, i.e. the so called Kodaira-Spencer
gravity \cite{Bershadsky:1993cx}, and the K\"ahler gravity
\cite{Bershadsky:1994sr} respectively. Further, in four dimensions there is
Pleba\'nski gravity \cite{Plebanski:1977zz} and in three dimensions
Chern-Simons gravity
\cite{Achucarro:1987vz,Witten:1988hc,Carlip:1998uc}.

The action principle for topological M-theory is Hitchin's
functional defined as the integration of the volume form in seven
dimensions, constructed with the $G_2$-holonomy invariant forms or
calibrations \cite{Hitchin:2001rw}. Volume functionals in six
dimensions also have been constructed by Hitchin
\cite{Hitchin:2000jd}. These actions reflect the possibility of
constructing volumes in terms of the K\"ahler structure or the
complex structure of the underlying six-manifold. In
\cite{Dijkgraaf:2004te} it was showed that these six-dimensional
Hitchin's functionals are related to the topological string theories
of A and B types. Also topological M-theory has been proposed as a
master system to derive all the form gravity theories in dimensions
lower or equal to seven \cite{Dijkgraaf:2004te}. This theory has
motivated recent work in exploring some new relations between
Hitchin's functionals in seven and six-dimensions and form gravity
theories in four and three dimensions
\cite{Herfray:2016azk,Herfray:2016std,Krasnov:2016wvc}.

In the present paper we give first a general overview of the case of
non-Abelian dualities in various gravity theories developed by the
authors in Refs.
\cite{GarciaCompean:1997tw,GarciaCompean:1998qh,GarciaCompean:1998wn,GarciaCompean:1999kj,GarciaCompean:2000ds,GarciaCompean:2001jx}.
These works were done motivated from Refs.
\cite{Ganor:1995em,Mohammedi:1995gy,Lozano:1995aq,Kehagias:1995ic},
which were based on the Rocek and Verlinde duality given in
\cite{Rocek:1991ps} and in the Buscher duality algorithm
\cite{Buscher:1987sk,Buscher:1987qj}.

A version of linearized gravitational duality was proposed in Ref.
\cite{Nieto:1999pn} based on our results mentioned in the previous paragraph.
Recently this subject has been intensively studied with very
interesting results, see for instance,
\cite{Bunster:2012km,Bunster:2014fca} and references therein. It is
not our purpose to overview this subject here.

Finally, as a new development in this direction, we study the
gravitational duality of Hitchin's functional in seven dimensions
and its relationship to the gravitational duality algorithm in the
corresponding six-dimensional theories.

In section 2 we give a brief review on the work performed by our
group regarding non-Abelian gravitational duality. In section 3 we
focus on the 3 dimensional Chern-Simons gravity where we describe in
detail the duality algorithm. Section 4 is devoted to study the
gravitational duality in the topological M-theory
\cite{Dijkgraaf:2004te}, which is described by using the Hitchin's
functionals \cite{Hitchin:2001rw} in seven-dimensional manifold with
$G_2$-holonomy. Further, the six-dimensional Hitchin's functionals
\cite{Hitchin:2000jd} are discussed. We find the corresponding dual
actions by using the duality algorithm. Moreover, using a relation
of Hitchin's functionals in six and seven dimensions, given via the
Hamiltonian flow, we prove that the duality in the six-dimensional
model is obtained from the duality in the seven dimensional theory.
Thus the parent action in the six-dimensional theory follows from
the corresponding action in the seven-dimensional case. Finally in
section 5 we give our final remarks.

\section{Gravitational duality in Form Theories of Gravity}

In this section we overview the idea of gravitational duality. There
are many conceptions of this duality. We are going to give a path to
our work. We will not intend to give a historical and detailed view
on the subject here. For a review about work done on the different
visions of gravitational duality and the development in early
stages, see \cite{Nurmagambetov:2006vz,Dehouck:2011xt}.

First of all, we review some aspects from Ref.
\cite{GarciaCompean:1997tw}. This work was inspired by the work in
Ref. \cite{Ganor:1995em}, where a dual action for non-supersymmetric
pure Yang-Mills theories is given. We start from the partition
function
\begin{equation}\label{parent1}
Z=\int{\cal D}A{\cal D}Ge^{i\int L_P d^Dx}=\int{\cal D}A{\cal
D}Ge^{i\int \left[g
G_a^{\mu\nu}G^a_{\mu\nu}+G_a^{\mu\nu}F^a_{\mu\nu}(A)\right] d^Dx},
\end{equation}
where $F_{\mu\nu}^a(A)=\partial_\mu A_\nu^a-\partial_\nu A_\mu^a+g
f_{bc}^{~~a}A_\mu^bA_\nu^c$. After integration of the auxiliary
field $G$ one get the original Yang-Mills Lagrangian
$L_{YM}=-\frac{1}{4 g}F_a^{\mu\nu}(A)F^a_{\mu\nu}(A)$. In order to
get the dual action, one can integrate with respect to the
gauge field $A$, from which follows
\cite{Ganor:1995em}
\begin{equation}\label{parent2}
Z=\int{\cal D}G\sqrt{\det(2g M)^{-1}}e^{i\int
\left(\frac{2\pi}{g}M^{-1ab}_{\mu\nu}\partial_\rho
G^{\rho\mu}_a\partial_\sigma G^{\sigma\nu}_b+g G^{\mu\nu}_a
G_{\mu\nu}^a\right) d^Dx},
\end{equation}
where $M_{ab}^{\mu\nu}=f^{~~c}_{ab}G_c^{\mu\nu}$. This is an action
of the Freedman-Townsend type \cite{Freedman:1980us}. Note that in
these computations it was not necessary to relate $G^{\mu\nu}_a$ to
$G_{\mu\nu}^a$ by any specific metric, the only condition is
invariance under the required symmetries. This formulation can be
generalized for a Lagrangian $L(\cal{F})$, which could also depend
on other fields. Consider the associated partition function
\begin{equation}\label{parent3}
Z=\int{\cal D}A{\cal D}G{\cal DF}e^{i\int\left\{L({\cal F})+2\pi
G_a^{\mu\nu}\left[{\cal
F}^a_{\mu\nu}-F^a_{\mu\nu}(A)\right]\right\}d^D x} .
\end{equation}
Proceeding as in the case of (\ref{parent1}), after integration of
$A$ the partition function becomes
\begin{equation}\label{parent3}
Z=\int{\cal D}G{\cal DF}\sqrt{\det(2 M)^{-1}}e^{i\int \left[2\pi
M^{-1ab}_{\mu\nu}\partial_\rho G^{\rho\mu}_a\partial_\sigma
G^{\sigma\nu}_b+G^{\mu\nu}_a {\cal F}_{\mu\nu}^a+L({\cal F})\right]
d^Dx}.
\end{equation}
For example, if $L({\cal
F})=\frac{\theta}{2\pi}F^{\mu\nu}_aF_{\mu\nu}^a$, where
$F^{\mu\nu}_a=\mathfrak{g}^{\mu\nu\rho\sigma}_{ab}F_{\rho\sigma}^b$,
is defined through some metric $\mathfrak{g}$. After integration of
${\cal F}$ the partition function becomes
\begin{equation}\label{parent4}
Z=\int{\cal D}Ge^{i\int \left(\frac{2\pi}{g}
M^{-1ab}_{\mu\nu}\partial_\rho G^{\rho\mu}_a\partial_\sigma
G^{\sigma\nu}_b-\frac{\pi}{2\theta} G^{\mu\nu}_a G_{\mu\nu}^a\right)
d^Dx}.
\end{equation}

This formulation has been applied in \cite{GarciaCompean:1997tw} to
topological gravity, considering the Pontryagin and Euler
topological invariants $L_P=\frac{\theta_G^P}{2\pi}\int
\varepsilon^{\mu\nu\rho\sigma}\delta_{ab}^{cd}R_{\mu\nu}^{~~ab}(\omega)R_{\rho\sigma
cd}(\omega)$ and $L_E=\frac{\theta_G^E}{2\pi}\int
\varepsilon^{\mu\nu\rho\sigma}\varepsilon_{abcd}R_{\mu\nu}^{~~ab}(\omega)R_{\rho\sigma}^{~~cd}(\omega)$,
where $a,b,c,d$ are Minkowski indices, $\omega_\mu^{ab}$ is the
spin connection and
$\delta_{ab}^{cd}=\frac{1}{2}(\delta_a^c\delta_b^d-\delta_a^d\delta_b^c)$.
Thus the gauge group is $SO(3,1)$. In this case in  (\ref{parent4})
we take
$G^{\mu\nu}_{ab}=\frac{1}{2}\varepsilon^{\mu\nu\rho\sigma}\eta_{ab,cd}G_{\rho\sigma}^{cd}$,
respectively
$G^{\mu\nu}_{ab}=\frac{1}{4}\varepsilon^{\mu\nu\rho\sigma}\varepsilon_{abcd}G_{\rho\sigma}^{cd}$,
where
$\eta_{ab,cd}=\frac{1}{2}(\eta_{ac}\eta_{bd}-\eta_{ad}\eta_{bc})$.
In fact, the equivalence of  $SO(3,1)$ with $SL(2,{\bf C})\times
SL(2,{\bf C})$ leads to the decomposition into self-dual and
anti-self-dual metrics
$\delta_{ab}^{cd}=\Pi^{(+)\,cd}_{~~ab}+\Pi^{(-)\,cd}_{~~ab}$ and
$\varepsilon_{ab}^{~~cd}=2i(\Pi^{(+)\,cd}_{~~ab}-\Pi^{(-)\,cd}_{~~ab})$.
Hence it is enough to consider
$L^{(\pm)}=\frac{\theta_G^{(\pm)}}{2\pi}\int
\varepsilon^{\mu\nu\rho\sigma}R_{~\mu\nu}^{(\pm)\,ab}(\omega)R^{(\pm)}_{\rho\sigma
ab}(\omega)$, where $G_{~\mu\nu}^{(\pm)\,ab}={\Pi^{(\pm)\,ab}}_{cd}G_{~\mu\nu}^{(\pm)\,cd}$ and
$G^{(\pm)\,\mu\nu}_{~~ab}=\frac{1}{2}\varepsilon^{\mu\nu\rho\sigma}\eta_{ac}\eta_{bd}G_{~\rho\sigma}^{(\pm)\,cd}
$.

For the MacDowell-Mansouri theory worked out in Ref.
\cite{GarciaCompean:1998qh} we have considered the classically
equivalent formulation $S=\int
d^4x\varepsilon^{\mu\nu\rho\sigma}\varepsilon_{abcd}
\left(\tau^{+}F_{\mu\nu}^{+\,ab}F_{\rho\sigma}^{+\,cd}-
\tau^{-}F_{\mu\nu}^{-\,ab}F_{\rho\sigma}^{-\,cd}\right)$, hence it
is sufficient to consider the self-dual action
$$
S=\int d^4x\varepsilon^{\mu\nu\rho\sigma}\varepsilon_{abcd}
F_{\mu\nu}^{+\,ab}(\omega,e)F_{\rho\sigma}^{+\,cd}(\omega,e)
$$
\begin{equation}
=2i\int
d^4x\varepsilon^{\mu\nu\rho\sigma}F_{\mu\nu}^{+\,ab}(\omega,e)F_{\rho\sigma
ab}^{+}(\omega,e),
\end{equation}
where
$F_{\mu\nu}^{\,ab}(\omega,e)=\partial_\mu\omega_\nu^{~ab}-\partial_\nu\omega_\mu^{~ab}+\omega_{\mu
a}^{~~c}\omega_{\nu
c}^{~~b}-\lambda^2(e_\mu^{~a}e_\nu^{~b}-e_\nu^{~b}e_\mu^{~a})$ is
the $ISO(1,3)$ field strength. In this case the partition function
is given by
\begin{equation}\label{zmm}
Z=\int{\cal D}\omega^{+}{\cal D} e{\cal D}G^{+}{\cal D}{\cal
F}^{+}e^{-2\int \left\{\frac{1}{\kappa}*{\cal
F}_{~~ab}^{+\mu\nu}{\cal F}^{+ab}_{~~\mu\nu}+2\pi
*G_{~~ab}^{+\mu\nu}[{\cal
F}^{+ab}_{~~\mu\nu}-F^{+ab}_{~~\mu\nu}(\omega,e)]\right\}d^4 x},
\end{equation}
where the $*$ corresponds to the Hodge dual. In this expression, the integration must be made considering the redundancy of the
self-dual integration variables, as can be seen from the
following identities: $u^{+ab}v_{ab}^{+}=-4u_{0i}^{+}v_{0i}^{+}$\,,
$*G_{ab}^{+\mu\nu}\omega_{\mu}^{+~ac}\omega_{\nu
c}^{+~b}=4*G_{ij}^{+\mu\nu}\omega_{0i\mu}^{+}\omega_{0j\nu}^{+}$ and
$*G_{ab}^{+\mu\nu}e_\mu^{~a}e_\nu^{~b}=*G_{ij}^{+\mu\nu}e_\mu^{~i}e_\nu^{~j}-i\varepsilon_{ijk}*G_{ij}^{+\mu\nu}e_\mu^{~0}e_\nu^{~k}$.
Thus, integrating successively the quadratic terms in ${\cal
F}_{\mu\nu}^{+0i}$, $\omega_\mu^{+0i}$ and $e_\mu^{~i}$, we get
\begin{align}
Z=\int{\cal D}G^{+}{\cal D}e\det\left({G^{+}}^{-1}\right)
e^{-8\int \left\{-\pi^2\kappa *G_{0i}^{+\mu\nu}G_{\mu\nu
0i}^{+}-\frac{\pi}{4}{G^{+}}_{\mu\nu}^{-1ij}\partial_\rho*G^{\sigma\mu}_{0i}\partial_\sigma*G^{\rho\nu}_{0j}+\frac{1}{16\pi\lambda^2}G^{(0)\mu\nu}e_{\mu0}e_{\nu0}\right\}d^4
x},
\end{align}
where the integration over ${\cal D} e$ stands for integration over
$e_\mu^{~0}$, and
$G^{(0)\mu\nu}={G^{+}}_{\rho\sigma}^{-1ij}\varepsilon_{ikl}\varepsilon_{jmn}*G^{+\rho\mu}_{~kl}*G^{+\sigma\nu}_{~mn}$.
After integration of $e_\mu^{~0}$, it follows the dual partition
function of (\ref{zmm})
\begin{align}
Z=\int{\cal D}G^{+}\det\left({G^{+}}^{-1}\right)\sqrt{\det\left({G^{(0)}}^{-1}\right)}
e^{-2\pi\int
\left\{{G^{+}}_{\mu\nu}^{-1ij}\partial_\rho*G^{\sigma\mu}_{0i}\partial_\sigma*G^{\rho\nu}_{0j}+4\pi\kappa
*G_{0i}^{+\mu\nu}G_{\mu\nu 0i}^{+}\right\}d^4 x}.
\end{align}
The MacDowell-Mansouri gravity can be generalized to supergravity.
This is done simply by promoting the gauge group SO(3,2) to the
supergroup OSp$(1|4)$. This theory was also studied in the context
of gravitational duality and a parent action was constructed and
its dual action was given in \cite{GarciaCompean:1998wn}.

For 3D Chern-Simons theories this approach simplifies considerably. In
this case we have $L_{CS}=\Tr (AH(A))$, where $H(A)$ is a $(D-1)$
form. We exemplify it for $D=3$, and following
\cite{GarciaCompean:1999kj}, consider the parent action
$I_P=i\tr\int\left(\alpha A{\cal F}+\beta G{\cal F}+\gamma GdA+\eta
GA^2\right)$, i.e.
\begin{equation}\label{pcs1}
I_P=i\int\varepsilon^{ijk}\left(\alpha A_i^a{\cal
F}_{jka}+\beta G_i^a{\cal F}_{jka}+\gamma G_i^a\partial_j
A_{ka}+\frac{\eta}{2}G_i^a f_{bca}A_j^bA_k^c\right),
\end{equation}
where
$\alpha$, $\beta$, $\gamma$ and $\eta$ are constants. We show that
independently of the integration order in the partition function,
the resulting dual action is the same (self-dual). In the next
section we make it explicitly since it is an important example in
order to compare with the dual actions of Hitchin's functionals in
seven and six dimensions (see section 4).

If we integrate first over $G$ in the partition function
\begin{equation}
Z=\int{\cal D}A{\cal D}G{\cal DF}\exp (iI_P), \label{partition}
\end{equation}
we get
\begin{equation}\label{daf}
Z=\int{\cal D}A{\cal D}{\cal F}\,\delta\left[\beta {\cal F}_{ij}^a+\frac{\gamma}{2}(\partial_i A_j^a-\partial_j A_i^a)+\frac{\eta}{2}f_{bca}A_i^b
A_j^c\right]e^{i\int\varepsilon^{ijk}\alpha A_i^a{\cal
F}_{jka} d^3x},
\end{equation}
from which, after integration over ${\cal F}$ gives
\begin{equation}\label{daf}
Z=\int{\cal
D}A\det\left(\frac{2\pi}{\beta}\right)e^{-i\frac{\alpha\gamma}{\beta}\int\varepsilon^{ijk}A_i^a\left(\partial_jA_{ka}+\frac{\eta}{2\gamma}
f_{bca}A_j^bA_k^c\right)d^3x}.
\end{equation}
Now, if we integrate over ${\cal F}$ in (\ref{partition}), we get
\begin{equation}\label{intgf}
Z=\int{\cal D}A{\cal D}G\delta(\alpha A_i^a+\beta
G_i^a)e^{i\int\varepsilon^{ijk}\left(\gamma
G_i^a\partial_jA_{ka}+\frac{\eta}{2}G_i^a
f_{bca}A_j^bA_k^c\right)d^3x}.
\end{equation}
Integrating now over $A$ and after the change $G^a_i \to - G^a_i$ it
yields
\begin{equation}
Z=\int{\cal
D}G\det\left(\frac{2\pi}{\alpha}\right)e^{-i\frac{\beta\gamma}{\alpha}\int\varepsilon^{ijk}G_i^a\left(\partial_jG_{ka}+
\frac{\beta\eta}{2\alpha\gamma}
f_{bca}G_j^bG_k^c\right)d^3x},\label{csg}
\end{equation}
and, otherwise, integrating over $G$ gives (\ref{daf}).

Finally, the integration over
$A$ and then over ${\cal F}$ can be performed if we observe that the
parent action (\ref{pcs1}) can be rewritten, after partial integration for the third term and simple algebraic manipulations, as
\begin{equation}\label{csff}
I_P=\int\left[\frac{\eta}{2}M_{ab}^{ij}\widetilde{A}_i^a\widetilde{A}_j^b-\frac{2\alpha^2}{\eta}M_{ij}^{-1ab}\widetilde{\cal
F}_a^i\widetilde{\cal
F}_b^j-\frac{\beta\gamma}{\alpha}\varepsilon^{ijk}G_i^a\left(\partial_jG_{ka}-\frac{\beta\eta}{2\alpha\gamma}
f_{bca}G_j^bG_k^c\right)\right]d^3x,
\end{equation}
where $M_{ab}^{ij}=\varepsilon^{ijk} f_{abc}G_k^c$,
$\widetilde{A}_i^a=A_i^a+\frac{2\alpha}{\eta}M_{ij}^{-1ab}\left({\cal
F}_b^j+\frac{\gamma}{2\alpha}\varepsilon^{jkl}\partial_kG_{lb}\right)$,
and $\widetilde{\cal
F}_a^i=F_a^i+\frac{\gamma}{2\alpha}\varepsilon^{ijk}\left(\partial_jG_{ka}+\frac{\beta\eta}{\alpha\gamma}f_{abc}G_j^bG_k^c\right)$.
Thus, the integration over $A$ and ${\cal F}$ in the partition function (\ref{partition}), leads to the integration of the first two gaussian terms in  the partition function of (\ref{csff}), whose contributions cancel up to a factor, following (\ref{csg}).

Therefore, the parent action (\ref{pcs1}) leads to the dual partition functions (\ref{daf}) and  (\ref{csg}). The dependence of the coupling constants in the determinants in these partition functions can be eliminated by rescalings, after which both actions coincide with
\begin{equation}
Z=\int{\cal D}Ae^{-i\alpha\beta\gamma\int\varepsilon^{ijk}A_i^a\left(\partial_jA_{ka}+\frac{\beta\eta}{2\gamma}
f_{bca}A_j^bA_k^c\right)d^3x},\label{csg}
\end{equation}
where constant factors have been discarded and for (\ref{csg}) a parity transformation $G\rightarrow -G$ must be made. Note that apparently the partition functions (\ref{daf}) and  (\ref{csg}) have inverted coupling constants, but after the preceding rescalings, they have the same dependence on the coupling constants. This is a characteristic of 3 dimensional Chern-Simons.

Gravity in three dimensions is also described as a gauge theory.
This is a Chern-Simons theory with gauge group being $SO(2,2)$,
$ISO(1,3)$ and $SO(1,3)$ according if the cosmological constant
takes negative, zero o positive values, respectively. These theories
also do admit a dual gravitational description and this was
described in Ref. \cite{GarciaCompean:1999kj}. This will be reviewed
in the next section. The generalization to Chern-Simons supergravity
is also obtained by promoting for instance the group SO$(2,2)$ to
the supergroup Osp$(2,2|1)$. The analysis of the gravitational
duality was performed in \cite{GarciaCompean:2000ds}.

Finally for the Pleba\'nski formulation, which is also known as of
the BF-type, it is somewhat different. In order to analyze it, we
consider the original complex version with a $SU(2)$ symmetry and an
action \cite{Plebanski:1977zz} $I=\int
\varepsilon^{\mu\nu\rho\sigma}\left[\frac{1}{\kappa}\Sigma_{\mu\nu}^iF_{\rho\sigma
i}(\omega)+\phi_{ij}\Sigma_{\mu\nu}^i\Sigma_{\rho\sigma}^j\right]d^4x$,
where $\phi$ is a traceless Lagrange multiplier matrix. The solution
of the constraints
$\varepsilon^{\mu\nu\rho\sigma}(\Sigma_{\mu\nu}^i\Sigma_{\rho\sigma}^j-\frac{1}{3}\delta^{ij}
\Sigma_{\mu\nu}^k\Sigma_{\rho\sigma k}) =0$ is
$\Sigma_{\mu\nu}^i=\frac{1}{2}(e_\mu^{~0}e_\nu^{~i}+\frac{i}{2}\varepsilon^i_{~jk}
e_\mu^{~j}e_\nu^{~k})$, which substituted into the action gives the
Palatini action. Following the previous analysis, we consider the
parent action
\begin{equation}\label{parentpl}
I=\int
\varepsilon^{\mu\nu\rho\sigma}\left(\alpha\Sigma_{\mu\nu}^i{\cal
F}_{\rho\sigma i}+\beta G_{\mu\nu}^i{\cal F}_{\rho\sigma i}+\gamma
G_{\mu\nu}^i\partial_\rho\omega_{\sigma i}+\eta
G_{\mu\nu}^i\varepsilon_{ijk}\omega_\rho^j\omega_\sigma^k+\phi_{ij}\Sigma_{\mu\nu}^i\Sigma_{\rho\sigma}^j\right).
\end{equation}
First we observe that the quadratic form in $\Sigma$, the last term,
cannot be integrated in combination with the first term to give a
${\cal F}^2$ term, because $\phi$ is traceless, and the first term
is contracted with $\delta_{ij}$. Furthermore, similar to the
Chern-Simons case, the integrations over ${\cal F}$ and $G$, and
over ${\cal F}$ and $\Sigma$, give all of them the Pleba\'nski
action
\begin{equation}
Z=\int{\cal D}\Sigma{\cal D}\omega{\cal D}\phi
e^{i\int\varepsilon^{\mu\nu\rho\sigma}\left[-\frac{\alpha\gamma}{\beta}\Sigma_{\mu\nu}^i\left(\partial_\rho\omega_{\sigma
i}+\frac{\eta}{\gamma}\varepsilon_{ijk}\omega_\rho^j\omega_\sigma^k\right)+\phi_{ij}\Sigma_{\mu\nu}^i\Sigma_{\rho\sigma}^j\right]}.
\end{equation}
However, the integration over $\omega$, ${\cal F}$ and $G$ in
(\ref{parentpl}) gives
\begin{equation}
Z=\int{\cal D}\Sigma{\cal
D}\phi\sqrt{\det\left(M^{-1}\right)}
e^{i\int\left(\frac{\alpha\gamma2}{\beta\eta}M_{\mu\nu}^{-1ij}\partial_\rho*\Sigma_i^{\rho\mu}\partial_\sigma*\Sigma_j^{\sigma\nu}+\varepsilon^{\mu\nu\rho\sigma}\phi_{ij}\Sigma_{\mu\nu}^i\Sigma_{\rho\sigma}^j\right)},
\end{equation}
where $M_{ij}^{\mu\nu}=\varepsilon_{ij}^{~~k}*\Sigma^{\mu\nu}_k$.

\section{Gravitational duality in 3D Chern-Simons gravity}

In the present section we explain the details of gravitational
duality for the Chern-Simons theory in three dimensions. This
subject has been discussed in the previous section for a general  Chern-Simons theory. However
here we intend to exhibit the details corresponding to gravitational Chern-Simons.

$2+1$ gravity dimensions is a theory that has played a very
important role as a toy model of four dimensional general relativity
at the classical and quantum levels. In \cite{Witten:1988hc} $2+1$
gravity was described in terms of the standard and exotic actions.
In \cite{GarciaCompean:2000ds} we showed that both actions
correspond to the self-dual and anti-self-dual of the Chern-Simons
actions with respect the Lorentz gauge group. The given Lie algebras
${\bf g}$ of the gauge groups have three generators constructed with
four capital Latin letters $A,B,C,D=0, \dots ,3$ and correspond to
${\bf g} = {\bf so}(3,1)$ and ${\bf g} = {\bf so}(2,2)$ for $\lambda
>0$ and $\lambda <0$ respectively.  The Lie algebra ${\bf g}$ is
generated by $M_{AB}$, which satisfy $[M_{AB},M_{CD}]=if_{\ \
ABCD}^{EF}M_{EF}$, where $f_{\ \ ABCD}^{EF}$ are its corresponding
structure constants.

Consider the non-Abelian Chern-Simons action
\begin{equation}
I(A)=\int_{X}{\frac{g}{4\pi }}{\rm Tr}(A\wedge H),
\end{equation}
where $g$ is the Chern-Simons coupling, $H=dA+{\frac{2}{3}}A\wedge
A$ and ${\rm Tr}$ is a quadratic form of ${\bf g}$ such that it
satisfies ${\rm Tr}(M_{AB}M_{CD})=\eta _{(AC}\eta _{BD)}$. The gauge
field $A=A_{i}^{AB}M_{AB}dx^{i}$ and the $H$-field
$H=H_{ij}^{AB}M_{AB}dx^{i}\wedge dx^{j}$. In local coordinates we
have
\begin{equation}
I(A)=\int_{X}d^{3}x{\frac{g}{4\pi }}\varepsilon
^{ijk}A_{i}^{AB}\bigg(\partial
_{j}A_{kAB}+{\frac{1}{3}}f_{ABCDEF}A_{j}^{CD}A_{k}^{EF}\bigg).
\label{cs}
\end{equation}
The duality algorithm require to propose a {\it parent action}, which in \cite{GarciaCompean:2000ds} has been proposed as
\begin{equation}
I_P(A,B,G) = \int_X d^3 x \varepsilon^{ijk} \bigg( a B^{AB}_i
H_{jkAB} + b A^{AB}_i G_{jkAB} + c B^{AB}_i G_{jkAB} \bigg),
\label{parent}
\end{equation}
where
\begin{equation}
H_{jkAB}= \partial_j A_{kAB} + {\frac{1 }{3}}
f_{ABCDEF}A^{CD}_jA^{EF}_k. \label{ache}
\end{equation}
The duality algorithm allows to recover the original action
(\ref{cs}) after an integration with respect the Lagrange
multipliers $B$ and $G$. In fact, in the preceding section we have
presented a somewhat more general parent action, although the
details of the computations in \cite{GarciaCompean:1999kj} are the
same, in such a way that the resulting actions, with $a=-g/4\pi$,
$b=1$ and $c=1$, are
\begin{equation}
{I}(A)=\int_{X}d^{3}x \frac{g}{4\pi}\varepsilon
^{ijk}A_{i}^{AB}\left(\partial _{j}A_{kAB}+{\frac{1}{3}}
f_{ABCDEF}A_{j}^{CD}A_{k}^{EF}\right).
\end{equation}
whereas the dual one is
\begin{equation}
I_D(B) = \int_X d^3 x \frac{g}{4\pi}\varepsilon
^{ijk}B_{i}^{AB}\bigg(\partial _{j}B_{kAB}+{\frac{1}{3}}
f_{ABCDEF}B_{j}^{CD}B_{k}^{EF}\bigg). \label{dual}
\end{equation}
In Chern-Simons gravity Newton's
gravitational constant $G_N$ is related to the Chern-Simons coupling
constant $g$ \cite{Carlip:1998uc} i.e.
\begin{equation}
g= - {1\over 4 G_N}.
\end{equation}
In the context of Abelian Chern-Simons theory, this duality was
previously working out by Balchandran \cite{Balachandran:1995dv}. In
particular, it was shown that in the Abelian case, the consistency
of the dual theory requires from periodic boundary conditions for
the dual fields $B_i^{AB}$. More recently this duality symmetry has
been explored in the context of supersymmetric Chern-Simons QED
\cite{Kapustin:1999ha}. Finally a more exhaustive analysis was
carried out in Ref. \cite{Witten:2003ya}.

\section{Gravitational Duality in Hitchin's gravity theories}
\label{sec:REV}

In this section we will discuss the so called topological M-theory
\cite{Dijkgraaf:2004te}. This is a theory in a seven-dimensional
manifold $X$ with $G_2$-holonomy and stable real 3 and 4-forms
$\Phi$ and $G = \star \Phi$.  The Hitchin's functionals
\cite{Hitchin:2001rw,Hitchin:2000jd} describing the volume of the
seven-manifold $X$ in terms of the stable forms are given by
\begin{equation}
V_7(\Phi) = \int_X \Phi \wedge \star \Phi, \label{one}
\end{equation}
and
\begin{equation}
V_7(G) = \int_X G \wedge \star G.
\label{two}
\end{equation}
The critical points of these actions determine special geometric
structures on $X$. For instance the variation of Eq. (\ref{one})
determines a metric with $G_2$-holonomy which can be constructed
through solutions of equations of motion
\begin{equation}
d \Phi =0, \ \ \ \ \ d \star \Phi =0.
\label{eomforphi}
\end{equation}
Another form we can see that this is true is as follows. If one
takes $\Phi$ to be an exact form: $\Phi = d B$ with $B$ being a two
form on $X$. Then immediately one has $d \Phi =0$. Then the
variation of the volume (\ref{one}) is given by
\begin{equation}
\delta V_7(\Phi) = 2 \int_X \delta \Phi \wedge \star \Phi.
\end{equation}
Taking into account that $\Phi = dB$ we have
\begin{equation}
\delta V_7(\Phi) = -2 \int_X  \delta B \wedge d \star \Phi =0.
\end{equation}
Then we have that Eqs. (\ref{eomforphi}) are fulfilled.  Similarly
this procedure can ba carried out for Eq. (\ref{two}).

A similar procedure can be implemented for Eq. (\ref{two}). It is
easy to see that the corresponding equations of motion are
\begin{equation}
d G =0, \ \ \ \ \ d \star G =0.
\label{eomforG}
\end{equation}

For seven manifolds which are the global product $X= M \times I$,
where $I$ is the finite interval. The stable forms in the
7-dimensional theory induce stable forms in the 6-dimensional theory
through a Hamiltonian flow. This determines the form of $\Phi$ and
$\star \Phi$ in terms of the 6-dimensional stable real 3-form $\rho$
and 2-form $k$ \cite{Dijkgraaf:2004te}.  This is given by
\begin{equation}
\Phi= \rho(t) + k(t) \wedge dt, \ \ \ \ \ \ \  \star\Phi= \sigma +
\widehat{\rho} \wedge dt,
\label{floweqs}
\end{equation}
where $\sigma= {1 \over 2} k \wedge k$. In terms of stable forms
$\rho$ and $k$ the Hitchin's action is written as
\cite{Dijkgraaf:2004te}
$$
V_7(\Phi) = \int_M  \rho \wedge \widehat{\rho} + {1 \over 2} \int_M
k \wedge k \wedge k
$$
\begin{equation}
= 2V_H(\rho) + 3 V_S(\sigma),
\label{sixdimaction}
\end{equation}
where
\begin{equation}
V_H(\rho) = {1 \over 2} \int_M \rho \wedge \widehat{\rho}
\label{volumenH}
\end{equation}
and
\begin{equation}
V_S(\sigma) = {1 \over 6} \int_M k \wedge k \wedge k.
\label{volumenS}
\end{equation}
Variations of (\ref{volumenH}) and (\ref{volumenS}) with respect to
$\rho$ and $k$ respectively lead to the following equations of
motion
\begin{equation}
d \rho =0,  \ \ \ \ \ \ \ \ dk =0.
\end{equation}
The first equation implies the existence of a closed holomorphic and
invariant $(0,3)$ form $\Omega$ on $M$ with $d\Omega =0$ and the
existence of a K\"ahler form $k$ on $M$. That means that $M$ is a
Calabi-Yau manifold.

\subsection{Gravitational Duality in Topological M-theory}
In this subsection we will find the dual action to the Hitchin's
actions (\ref{one}) and (\ref{two}). In order to do that we consider
the following parent action
\begin{equation}
I_P(\Phi,B, \Lambda) = \int_X \bigg(a B \wedge \star \Phi + b \Phi
\wedge \star \Lambda + c B \wedge \star \Lambda \bigg),
\label{parent}
\end{equation}
where $\Phi$, $B$, $\Lambda$ belong to $\Omega^3(X)$ and $a$, $b$
and $c$ are undetermined constants. Integrating out with respect to
the Lagrange multipliers $\Lambda$ and $B$ we will regain the
original action (\ref{one}). Thus in Euclidean signature
\begin{equation}
\exp\big\{ - I(\Phi)\big\}= \int DB D \Lambda \exp\bigg( -
I_P(\Phi,B, \Lambda) \bigg),
\label{generalpartitionf}
\end{equation}
after integration we have
\begin{equation}
I(\Phi)= -{ab \over c} \int_X \Phi \wedge \star \Phi.
\end{equation}
If we select the constants , $b=1$ and $c=-1$, then
\begin{equation}
I(\Phi)= a  \int_X \Phi \wedge \star \Phi.
\label{almostoriginal}
\end{equation}
If we take $a=1$ then
\begin{equation}
I(\Phi)= V_7(\Phi).
\end{equation}

Now we can obtain the dual action by integrating out
(\ref{generalpartitionf}) with respect to $B$ and $\Phi$
\begin{equation}
\exp\big\{ - I_D(\Lambda)\big\}= \int D\Phi D B \exp\bigg( -
I_P(\Phi,B, \Lambda) \bigg).
\label{dualPF}
\end{equation}
Integrating out first with respect to $B$ we get
\begin{equation}
\int D\Phi D \Lambda \ \delta \big[ \star(a \Phi + c \Lambda) \big]
\exp\bigg( - b \int_X \Phi \wedge \star\Lambda \bigg).
\end{equation}
Further integration with respect to $\Phi$ leads to the dual action
\begin{equation}
{I_D}(\Lambda) = -{bc \over a} \int_X \Lambda \wedge \star \Lambda.
\end{equation}
As in the derivation of action (\ref{almostoriginal}) we have
\begin{equation}
{I_D}(\Lambda) = {1\over a} \int_X \Lambda \wedge \star \Lambda.
\end{equation}
Thus the dual action looks exactly of the same form as the Hitchin's
action (\ref{almostoriginal}) but with the coupling constant
inverted and interchanging the original degrees of freedom $\Phi$ by
the dual variables $\Lambda$, which are the Lagrange multipliers.

Once again if $a=1$ we have
\begin{equation}
{I_D}(\Lambda) = V_7(\Lambda),
\end{equation}
where
\begin{equation}
V_7(\Lambda) =\int_X \Lambda \wedge \star \Lambda.
\label{eomLambda}
\end{equation}
It is immediate to see that the equations of motion associated to
the dual action (\ref{eomLambda}) are
\begin{equation}
d \Lambda =0, \ \ \ \ \ \   d \star \Lambda =0.
\label{dualEOM}
\end{equation}
Then we conclude that the topological M-theory with action
(\ref{one}) is self-dual.

\subsection{Derivation of the parent action in six dimensions from Topological M-theory}

As we mentioned before action (\ref{one}) can be reduced to a theory
in six dimensions which is the linear combination given by
(\ref{sixdimaction}). Now we will show that under certain conditions
the duality algorithm in the seven dimensional theory can be induced
to a duality procedure in six dimensions from the action
(\ref{parent}).

In addition to Eqs. (\ref{floweqs}) we have
\begin{equation}
B = b_3 + b_2 \wedge dt, \ \ \ \ \ \   \star B = {b}_4 +
\widehat{b}_3 \wedge dt,
\end{equation}
and something similar happens for the Lagrange multiplier
\begin{equation}
\Lambda = \lambda_3 + \lambda_2 \wedge dt, \ \ \ \ \ \   \star
\Lambda = {\lambda}_4 + \widehat{\lambda}_3 \wedge dt.
\end{equation}
We impose that all of them satisfy the Calabi-Yau condition
\begin{equation}
b_3 \wedge b_2 =0, \ \ \ \ \  \lambda_3 \wedge \lambda_2 =0.
\label{CYcondition}
\end{equation}
Moreover we assume the same dependence for $\widehat{\rho}$ and
$\widetilde{\lambda}_3$. That is, if we have $\widehat{\rho}(\rho)
=\widetilde{\lambda}_3(\lambda_3)$, this implies that $\rho
=\lambda_3$. With these conditions it is possible to show that the
parent action (\ref{parent}) can be reduced to the linear
combination of two parent actions in six dimensions
\begin{equation}
I_P =  I_P(\rho,b_3,\widehat{\lambda}_3) + I_P(\sigma,b_2,
\widehat{\lambda}_4),
\end{equation}
where
\begin{equation}
I_P(\rho,b_3,\widehat{\lambda}_3) = a \int_M b_3 \wedge
\widehat{\rho} + b \int_M \widehat{\lambda}_3 \wedge \rho + c \int_M
\widehat{\lambda}_3 \wedge b_3,
\label{parentB}
\end{equation}
and
\begin{equation}
I_P(\sigma,b_2, \widehat{\lambda}_4) = a \int_M \sigma \wedge {b}_2
+ b \int_M \widehat{\lambda}_4 \wedge k + c \int_M
\widehat{\lambda}_4 \wedge b_2. \label{parentA}
\end{equation}

\subsection{Gravitational Duality in six dimensions}

Now we describe the duality in six dimensions. We start from the
action (\ref{parentB}). Integrating out with respect the Lagrange
multiplier $\widehat{\lambda}_3$ we go back to the original action
\begin{equation}
I(\rho)= -{ab \over c} \int_M \rho \wedge \widehat{\rho}.
\end{equation}
If we take $a=b=1$ and $c=-1$ as before. Then we get
\begin{equation}
I(\rho)= V_H(\rho).
\end{equation}

Now the dual action can be obtained by calculating the effective
action and integrating out with respect to $b_3$ and then with
respect to $\widehat{\rho}$. This is given by
\begin{equation}
\exp\big\{ - I_D(\widehat{\lambda}_3)\big\}=\int D
\widehat{\lambda}_3 Db_3 D \widehat{\rho} \exp\bigg( -
I_P(\rho,b_3,\widehat{\lambda}_3) \bigg).
\end{equation}
Integration with respect to $b_3$ yields
\begin{equation}
\int D \widehat{\lambda}_3  D \widehat{\rho} \  \delta(a
\widehat{\rho} + c \widehat{\lambda}_3)\exp\bigg( - b \int_M \rho
\wedge \widehat{\lambda}_3 \bigg).
\end{equation}
Now we use the arguments given after Eq. (\ref{CYcondition}),
integration on $\widehat{\rho}$ and $\widehat{\lambda}_3$ can be
expressed as integrations with respect to $\rho$ and $\lambda_3$,
then
\begin{equation}
\int D \widehat{\lambda}_3  D {\rho} \  \delta(a {\rho} + c
{\lambda}_3)\exp\bigg( - b \int_M \rho \wedge \widehat{\lambda}_3
\bigg).
\end{equation}
Then integration with respect to ${\rho}$ determines the dual theory
\begin{equation}
I_D(\rho)= -{bc \over a} \int_M  {\lambda}_3 \wedge
\widehat{\lambda}_3.
\end{equation}
Again, for $b=1$ and $c=-1$ we have
\begin{equation}
I_D(\rho)= {1\over a} \int_M  {\lambda}_3 \wedge
\widehat{\lambda}_3. \label{dualRO}
\end{equation}
It is a self-dual theory that inverts the coupling $a$ and
interchanges the original degrees of freedom $\rho$ by $\lambda_3$.

We now discuss the duality coming from the parent action
(\ref{parentA}). Once again integration with respect to the Lagrange
multiplier $\widehat{\lambda}_4$ leads to the original action
\begin{equation} I(\sigma)= {a} \int_M \sigma \wedge k.
\end{equation}

Finally we will get the dual action. Before that we make the
assumption that the Lagrange multiplier $\widehat{\lambda}_4$ can be
rewritten as
\begin{equation}
\widehat{\lambda}_4 = {1 \over 2} \widetilde{\lambda} \wedge
\widetilde{\lambda}.
\label{conditionone}
\end{equation}
The dual action is then defined by
\begin{equation}
\exp\big\{ - I_D(\widetilde{\lambda})\big\}= \int Db_2 Dk \exp\big\{
- I_P(\sigma,b_2, \widetilde{\lambda})\big\}.
\end{equation}
Integration with respect to the field $b_2$ including condition
(\ref{conditionone}) leads to
\begin{equation}
\int D \widetilde{\lambda}  Dk  \ \delta(a k + c
\widetilde{\lambda})\exp\bigg( - {b \over 2}\int_M
\widetilde{\lambda} \wedge \widetilde{\lambda} \wedge k \bigg).
\end{equation}
Finally integration with respect to $k$ determines the dual theory
for $a=b=1$ and $c=-1$
\begin{equation}
I_D(\widetilde{\lambda})= {1 \over 2} \int_M \widetilde{\lambda}
\wedge \widetilde{\lambda} \wedge \widetilde{\lambda}.
\label{dualK}
\end{equation}

\section{Final Remarks}
\label{sec:final}

In the present article we give an overview of some of our results
regarding gravitational duality in some gravity theories
\cite{GarciaCompean:1997tw,GarciaCompean:1998qh,GarciaCompean:1998wn,GarciaCompean:1999kj,GarciaCompean:2000ds,GarciaCompean:2001jx}.
In these papers we found some explicit dual actions to some specific
theories of gravity. The duality procedure was implemented from the
Rocek-Verlinde non-Abelian duality algorithm applied to Yang-Mills
theories
\cite{Ganor:1995em,Mohammedi:1995gy,Lozano:1995aq,Kehagias:1995ic}.

In section 2, we reviewed the gravitation duality in a unified
framework that contains the cases of topological gravity,
MacDowell-Mansouri gravity, Chern-Simons gravity and BF-gravity. In
order to present a complete case, in section 3 we overview in more
detail the corresponding gravitational duality to Chern-Simons
gravity in 2+1 dimensions.

In section 4 we give a new contribution to the subject. We apply for
the first time the duality algorithm to the Hitchin's volume
functionals in seven and six dimensions. Hitchin's functional in
seven dimensions is the starting point to define topological
M-theory \cite{Dijkgraaf:2004te}. In the present article we find the
dual action which is written in terms of the dual degrees of freedom
$\Lambda$ and it is observed to be {\it self-dual} since it has the
same form as the original action. Furthermore the dual action has
inverted the coupling constant. Moreover the dual theory has the
same equations of motion (\ref{dualEOM}) than the original theory.
Thus they have the same dynamics and both theories are classically
equivalent.

The duality algorithm was also implemented for the Hitchin's
functionals in six dimensions. In this case we have two volume
functionals $V_H(\rho)$ and $V_S(\sigma)$. We have found the dual
actions for such functionals given by expressions (\ref{dualRO}) and
(\ref{dualK}). Moreover we showed that the corresponding parent
giving rise to these actions given by expressions (\ref{parentB})
and (\ref{parentA}) follows, under certain sensible conditions, from
the parent action of the topological M-theory (\ref{parent}). Thus
we find that the duality algorithm in six dimensions come from the
duality procedure of the underlying seven-dimensional theory. It is
expected that this connection can be carried out to other duality
procedures in form theories of gravity in lower dimensions. In a
future work we expect to report our results in the search of this
web of dualities from M-theories to lower dimensions including the
four and three dimensions. It would be interesting to investigate if
the recent results
\cite{Herfray:2016azk,Herfray:2016std,Krasnov:2016wvc}, will be of
some relevance in this analysis.


 \vspace{.5cm}
\centerline{\bf Acknowledgments} \vspace{.5cm}

HGC thanks the Universidad de Guanajuato and Prof. OO for
hospitality during the sabbatical stay. CR thanks VIEP-BUAP and
PFCE-SEP for the financial support. OO was supported by CONACyT:
Project 257919; Universidad de Guanajuato: Project CIIC130/2018 and
PRODEP.



\end{document}